\documentclass[
 reprint,
 amsmath,
 amssymb,
 aps,
 superscriptaddress,
 preprintnumbers,
 nofootinbib,
 longbibliography
]{revtex4-2}

\usepackage{hhline}
\usepackage{vmargin,epsfig}
\usepackage{graphicx}
\usepackage[english]{babel}
\usepackage[toc,page]{appendix}
\usepackage{parskip}
\usepackage{xspace}
\usepackage{braket}
\usepackage{mathtools}
\usepackage{multirow}
\usepackage{booktabs}
\usepackage{subfigure}

\usepackage[colorlinks=true]{hyperref}
\usepackage[dvipsnames]{xcolor}
\hypersetup{
colorlinks=true,
linkcolor=blue,
linktocpage=true,
citecolor=black,
urlcolor=blue
}

\usepackage{etoolbox}
\makeatletter
    \newcommand\newsubsupcommand[4]{\newcommand#1{#2\sc@subp{#3}{#4}}}
    \def\sc@subp#1#2{%
        \let\sc@subflag\undefinded%
        \let\sc@supflag\undefinded%
        \def\sc@thesub{#1}%
        \def\sc@thesup{#2}%
        \sc@proc%
    }%
    \def\sc@proc{%
        \@ifnextchar{_}{\def\sc@subflag{}\sc@mergesubs}{%
            \@ifnextchar{^}{\def\sc@supflag{}\sc@mergesups}{
                \ifdef{\sc@subflag}{}{_{\sc@thesub}}%
                \ifdef{\sc@supflag}{}{^{\sc@thesup}}%
            }%
        }%
    }%
    \def\sc@mergesubs#1#2{_{\sc@thesub#2}\sc@proc}%
    \def\sc@mergesups#1#2{^{\sc@thesup#2}\sc@proc}%
\makeatother

\newsubsupcommand{\phiL}{\varphi}{L}{}
\newsubsupcommand{\phiR}{\varphi}{R}{}
\newsubsupcommand{\psiL}{\psi}{L}{}
\newsubsupcommand{\psiR}{\psi}{R}{}

\newcommand{\fsl}[1]{\ensuremath{\mathrlap{\!\not{\phantom{#1}}}#1}}
\newcommand{\namedref}[2]{\hyperref[#2]{#1~\ref*{#2}}}

\makeatletter
\def\mr@ignsp#1 {\ifx\:#1\@empty\else #1\expandafter\mr@ignsp\fi}%
\newcommand{\multiref}[1]{\begingroup
\xdef\mr@no@sparg{\expandafter\mr@ignsp#1 \: }%
\def\mr@comma{}%
\@for\mr@refs:=\mr@no@sparg\do{\mr@comma\def\mr@comma{,\,}\ref{\mr@refs}}%
\endgroup}
\makeatother
\renewcommand{\eqref}[1]{(\multiref{#1})}

\newcommand{\be}{\begin{equation}}
\newcommand{\ee}{\end{equation}}
\newcommand{\bea}{\begin{eqnarray}}
\newcommand{\eea}{\end{eqnarray}}
\newcommand{\bei}{\begin{itemize}}
\newcommand{\eei}{\end{itemize}}

\definecolor{green1}{HTML}{244819}
\definecolor{cyan1}{HTML}{37cdaa}
\definecolor{blue1}{HTML}{5d7ac4}
\definecolor{red1}{HTML}{d0482a}
\definecolor{purple1}{HTML}{845ea8}
\definecolor{orange1}{HTML}{e07229}

\begin{document}

\allowdisplaybreaks


\title{Three-loop helicity amplitudes of four-lepton scattering in QED}

\newcommand{\higgs}{Higgs Centre for Theoretical Physics, School of Physics and Astronomy The University of Edinburgh, Edinburgh EH9 3FD, Scotland, UK}

\newcommand{\genericaffiliation}{University of University, University Road, County, Country}

\newcommand{\padova}{Dipartimento di Fisica e Astronomia, Universita di Padova, Via Marzolo 8, 35131 Padova, Italy}

\newcommand{\infn}{INFN, Sezione di Padova,
Via Marzolo 8, I-35131 Padova, Italy.}

\newcommand{\liverpool}{Department of Mathematical Sciences, University of Liverpool, Liverpool L69 3BX, 
United Kingdom
}

\author{Giulio Crisanti}
\affiliation{\higgs}

\author{Thomas Dave}
\affiliation{\liverpool}

\author{Pierpaolo Mastrolia}
\affiliation{\padova}
\affiliation{\infn}

\author{\\Jonathan Ronca}
\affiliation{\padova}
\affiliation{\infn}

\author{Sid~Smith}
\affiliation{\higgs}
\affiliation{\padova}
\affiliation{\infn}

\author{William J. Torres~Bobadilla}
\affiliation{\liverpool}


\begin{abstract}
We present the analytic expressions of the three-loop virtual corrections to the helicity amplitudes of $2\to2$ four-fermion scattering processes in massless QED. The contributing Feynman diagrams are grouped into integrand families characterised by independent Symanzik polynomials and decomposed in terms of master integrals using an optimised integration-by-parts strategy. Upon the renormalisation of the ultraviolet divergences and the extraction of the universal infrared pole structure, the finite results are expressed in terms of generalised polylogarithms up to transcendental weight six.
Amplitudes for dimuon production in electron-positron annihilations, electron-muon scattering, and Bhabha scattering are explicitly derived.

\end{abstract}

\maketitle

\section{Introduction}

Precision predictions for four-fermion processes in quantum electrodynamics (QED) play a central role in the physics program of modern lepton factories.
At flavour factories and low-energy precision experiments, including
BaBar~\cite{BaBar:2013agn},
Belle~II~\cite{Belle-II:2024vuc},
BESIII~\cite{BESIII:2024lbn},
Kloe-2~\cite{KLOE-2:2017fda},
MUonE~\cite{MUonE:2016hru,CarloniCalame:2015obs}, and
FCC-ee~\cite{FCC:2025lpp},
experimental uncertainties have reached, or aim to reach, the per mille or sub-per-mille level. In this regime, theoretical uncertainties from missing higher-order radiative corrections become a limiting factor, motivating calculations beyond the current state of the art.

The computation of multi-loop QED corrections to four-fermion scattering amplitudes is therefore essential to match the experimental precision in order to fully exploit the discovery potential of these facilities. In this context, three-loop corrections are expected to play an increasingly important role in reducing uncertainties associated with luminosity normalisations, as well as precision observables derived from leptonic final states. In particular Bhabha scattering, as well as the related $e^+ e^- \to \mu^+ \mu^-$ and $e \mu$ processes provide the theoretical normalisation for luminosity measurements and precision Standard Model tests~\cite{Aliberti:2024fpq}.

Four-lepton scattering in QED is a process that, at leading order, receives contributions from tree-level diagrams. The Born term \cite{berestetskii1956formation} and the next-to-leading corrections \cite{Berends:1973fd,Berends:1982dy,Jadach:1984hwn} have been known for a long time. The next-to-next-to-leading corrections were computed in \cite{Broggio:2022htr}, and required the evaluation of two-loop corrections to the scattering amplitude \cite{Banerjee:2020tdt,Bonciani:2021okt,Delto:2023kqv,Gerasimov:2025pqf}.

The three-loop virtual corrections, considered in this work for the first time, contribute to next-to-next-to-next-to-leading order predictions and are essential for several reasons. First, they reduce the residual renormalisation-scale dependence and stabilise perturbative predictions in kinematic regions relevant for luminosity monitoring and precision fits. Second, they control the interplay between purely leptonic corrections and vacuum-polarisation effects, which is crucial when extracting hadronic contributions from precision measurements of differential cross sections.

Beyond their phenomenological impact, three-loop computations in four-fermion QED processes are also of fundamental theoretical interest. They provide a ``laboratory'' for the development of modern multi-loop techniques. These include integral decomposition, analytic and numerical evaluation methods for master integrals, as well as the discover of novel analytic structures present in such multi-scale Feynman integrals. These developments are increasingly important not only for QED but also for mixed QCD–electroweak corrections and for precision calculations across the Standard Model.

\begin{figure}[t]
    {\centering
    \includegraphics[scale=0.22]{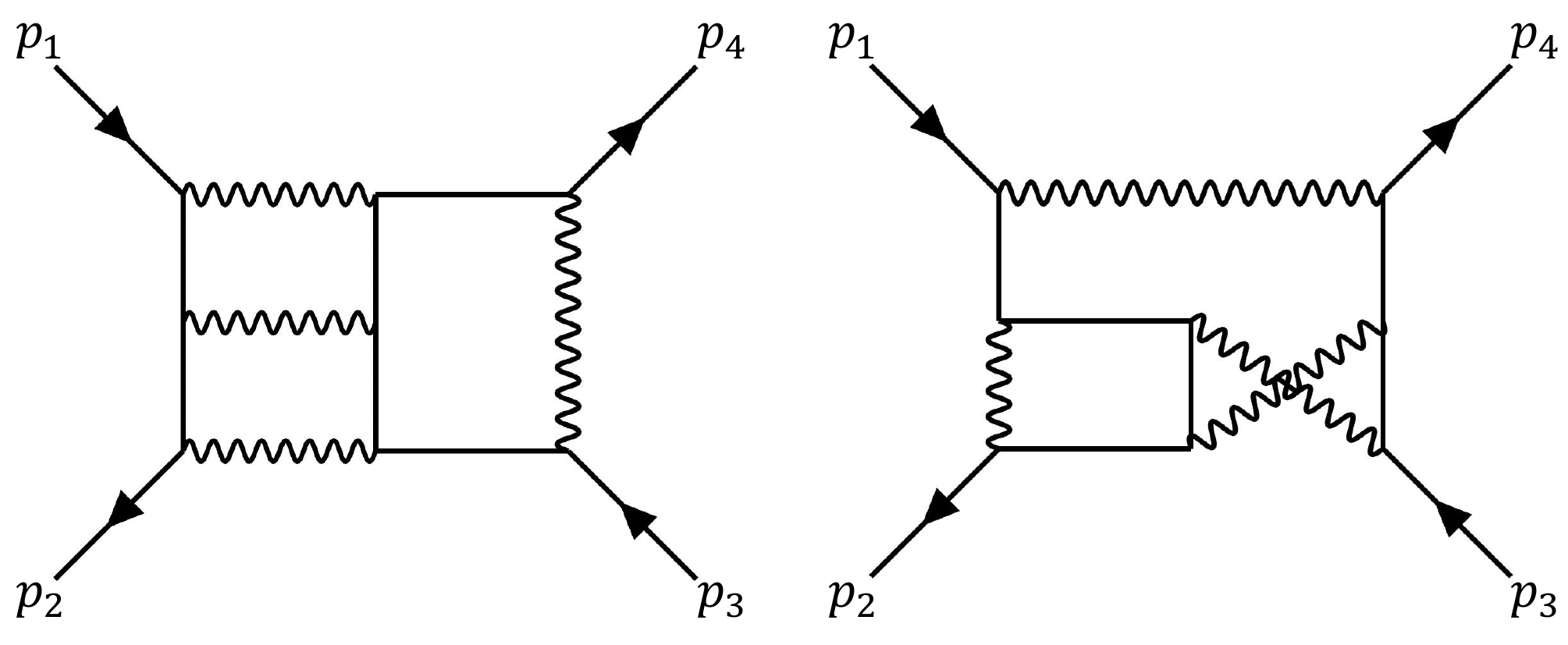}
    }
    \caption{Representative three-loop diagrams of four-fermion scattering in QED, 
    with outgoing external momenta $p_i$: solid lines represent leptons, and wavy lines, photons.
    }
       \label{fig:ordering}
\end{figure}

In this Letter we present analytical results for the non-vanishing helicity amplitudes relevant to four-fermion scattering processes in massless QED at the three-loop order. These results 
extend the virtual corrections considered in Refs.~\cite{Arbuzov:1995qd,Bern:2000ie,Dave:2024ewl} 
to one order higher in the QED perturbative expansion
and provide a key ingredient for next-generation precision predictions required by current and future lepton-scattering experiments.

The amplitudes are decomposed into four-dimensional Lorentz-covariant tensor structures~\cite{Chen:2019wyb,Peraro:2019cjj,Peraro:2020sfm}. The loop integrals entering the associated form factors are regulated using 
dimensional regularisation.
By employing integration-by-parts relations generated with an optimised decomposition algorithm, the form factors, and hence the scattering amplitudes, are reduced to a basis of master integrals~\cite{Henn:2020lye}, up to crossings of external kinematics~\cite{Bargiela:2021wuy}. 
These integrals are evaluated by combining known analytic results for a subset of kinematic configurations with analytic continuation across different physical regions~\cite{Moriello:2019yhu,Hidding:2020ytt}, supplemented by high-precision numerical evaluations and arithmetic reconstruction. 
Ultraviolet and infrared divergences are removed through UV renormalisation and IR subtraction respectively, using known lower-loop results. This procedure allows for the extraction of finite remainders. The final expressions for all helicity amplitudes are given in terms of generalised polylogarithms up to transcendental weight six.
\section{Scattering Amplitude}

We consider the scattering process:
\begin{equation}
    e^{-}(-p_{1})e^{+}(-p_{2})\rightarrow\mu^{+}(p_{3})\mu^{-}(p_{4})
    \label{eq:amplitude_mumu}
\end{equation} 
for outgoing massless fermions,  
$0 \to p_1+p_2+p_3+p_4$, with
$p_i^2=0$ ($i=1,\ldots,4$). 
Scattering amplitudes are expressed as a series expansion in the bare coupling constant $\alpha_{B}=e_B^2/4\pi$, which up to the third order reads
\begin{equation}
    \mathcal{A}_{B} = \sum_{L=0}^{3}\left(\frac{\alpha_{B}}{2\pi}\right)^{L}\mathcal{A}_{B}^{(L)}\,,
    \label{eq:bare_amp_decomposition}
\end{equation}
where $\mathcal{A}_{B}^{(L)}$ is the bare amplitude at $L$ loop order. 
The amplitude is a function of the  Mandelstam invariants, 
$s = (p_{1}+p_{2})^{2}\,,
        t = (p_{2}+p_{3})^{2}\,,
        u = (p_{1}+p_{3})^{2}\,,$ with $u= -s-t$.
The dimuon production region of the amplitude~\eqref{eq:amplitude_mumu} is given by
\begin{equation}
    s>0, \quad t,u<0\,.
\end{equation}
Because external momenta live in strictly four space-time dimensions, 
each component ${\cal A}^{(L)}_{B}$ admits a decomposition in terms of two independent four-dimensional gauge invariant tensor structures~\cite{Chen:2019wyb,Peraro:2019cjj,Peraro:2020sfm}: 
\begin{equation}
 \mathcal{A}_{B}^{(L)} = 
 \mathcal{F}^{(L)}_{1}
 \mathcal{T}_{1,\mu\mu}
 +\mathcal{F}^{(L)}_{2}\mathcal{T}_{2,\mu\mu}
 \,,
 \label{eq:eemmumu_tensor_decomposition}
\end{equation}
with
\begin{equation}
 \begin{aligned}
\label{eq:eemumu_Tensor_Structures}
 &\mathcal{T}_{1,\mu\mu}
=\overline{u}(p_2)\gamma_{\sigma} v(p_1)\times\overline{u}(p_4)\gamma^{\sigma} v(p_3)\,,&&\\
&
\mathcal{T}_{2,\mu\mu}
=\overline{u}(p_2)\fsl{p_3} v(p_1)\times\overline{u}(p_4)\fsl{p_1} v(p_3)\,,&&
\end{aligned}
\end{equation}

where the subscript ``$\mu\mu$'' indicates the final dimuon state.
The scalar \textit{form factors} $\mathcal{F}_{i}^{(L)}$ can be extracted by projection, according to the formula:
\begin{equation}
\begin{aligned}
    &\mathcal{F}^{(L)}_{i} = 
    \braket{\mathcal{T}_{i},\mathcal{T}_{j}}^{-1}\braket{\mathcal{A}^{(L)}_{B},\mathcal{T}_{j}}, \quad \\
    &{ \rm with \ }
    \braket{\mathcal{T}_{i}, \mathcal{T}_{j}} = \sum_{\text{spins}}\mathcal{T}_{i}\mathcal{T}_{j}^{\dagger}\,.
\end{aligned}
\end{equation}

In this letter, we evaluate the third-order virtual corrections term $\mathcal{A}_{B}^{(3)}$ of eq.~\eqref{eq:bare_amp_decomposition}
in dimensional regularisation,
 with $d=4-2\varepsilon$ space-time dimensions. 
These contributions are obtained from three-loop diagrams as those depicted in Fig.~\ref{fig:ordering}.

\subsection{Diagrams, Graphs and Integrals}

\begin{table}[t]
\begin{tabular}{cllll}
\toprule
        {Loops\phantom{w} }&	{$N_f^0$}&	{$N_f^1$}&	{$N_f^2$}&	{$N_f^3$} \\
        \midrule
        0 & 1 & -- & -- & -- \\
        1 & 4 & 1 & -- & -- \\
        2 & 35 & 11 & 1 & -- \\
        3 & 384\phantom{w} & 162\phantom{w} & 21\phantom{w} & 1 \\
\bottomrule
\end{tabular}
\caption{Number of diagrams up to three loops, collected according to the number of closed fermion loops.
}\label{tab:number_of_diagrams}
\end{table}

The number of Feynman diagrams contributing to the virtual corrections of 
$e^- e^+ \to \mu^+ \mu^-$ are listed in Tab.~\ref{tab:number_of_diagrams}, up to three loops.
Feynman diagrams were generated using {\sc FeynArts}~\cite{Hahn:2000kx} with {\sc FeynCalc}~\cite{Mertig:1990an,Shtabovenko:2016sxi,Shtabovenko_2025}, for a total of 1008 contributing graphs (distinguishing those with distinct fermion flavour loops). This number can be significantly reduced to 568
by considering the (fermion exchange) symmetry arising from the massless fermion approximation, and by eliminating those that vanish due to Furry's theorem.

\subsubsection{Graph mapping and integrals' relations}

 An in-house topology-mapping algorithm was employed to group the contributing diagrams into 158 equivalence classes, identified by their graph (Symanzik) polynomials, collected in two sets of 72 planar and 86 non-planar parent structures.

The generic form of the scalar integrals belonging to each topology   reads as:\begin{equation}
    I_{n_{1},\dots,n_{15}} = 
    e^{3\epsilon\gamma_\text{E}}
    (s)^{\frac{3d}{2}-n}
    \int\prod_{j=1}^{3}\frac{d^{d}\ell_{j}}{i\pi^{d/2}}\prod_{i=1}^{15}D_{i}^{-n_{i}}\,,
\label{eq:int_I}
\end{equation}
where we factor out the dependence on the kinematic scale $s$, $n=n_1+\hdots+n_{15}$,
and 
in terms of 15 generalised denominators $D_i$, out of which 10 are  loop denominators, 
while the remaining 5 are auxiliary denominators.

The master equation: 
\begin{equation}
    \int\prod_{j=1}^{3}\frac{d^{d}\ell_{j}}{i\pi^{d/2}}\frac{\partial}{\partial \ell_{k}^{\mu}}\left(q^{\mu}\prod_{i=1}^{15}D_{i}^{-n_{i}}\right)= 0  \, ,
    \label{eq:ibpequations}
\end{equation}
where $q\in\{\ell_1,\ell_2,\ell_3,p_1,p_2,p_3\}$, 
forms the basis for deriving integration-by-parts (IBP) identities~\cite{Tkachov:1981wb,Chetyrkin:1981qh,Laporta:2000dsw} among the dimensionally regulated integrals $I_{n_{1},\dots,n_{15}}$. 
By seeding the symbolic equation in Eq.\eqref{eq:ibpequations} with different choices of the indices $n_{i}$, one generates a large linear system of relations among these integrals. Solving this system yields the reduction of all integrals $I_{n_{1},\dots,n_{15}}$ to a basis of independent master integrals, with exact dependence on the spacetime dimension $d$ and on the kinematic variables $s$ and $t$.

The reduction of the integrals encountered in this calculation presents a significant computational challenge.
The most difficult contributions include diagrams with numerator rank as high as six and up to two powers of inverse propagators. Their decomposition was accomplished using a newly developed algorithm that integrates several complementary optimization strategies to efficiently construct and solve the corresponding IBP systems.

These included the use of {\it spanning cuts} \cite{Larsen:2015ped} to decompose the reduction procedure into smaller, more tractable subsystems. Furthermore, the number of integrals entering the IBP relations was systematically restricted through a combined application of {\it syzygy-based equations} \cite{Gluza:2010ws,Wu:2023upw,Wu:2025aeg,Smith:2025xes} and {\it improved seeding algorithms} \cite{Lange:2025fba,Bern:2024adl,Driesse:2024xad,Smirnov:2025prc} specifically tailored to the syzygy approach. A very similar procedure has recently been employed successfully in the IBP reduction of six-loop integrals arising in classical gravity computations \cite{Brunello:2025gpf}.

\begin{figure}[t]
\subfigure[26 MIs]{\includegraphics[page=1, scale=0.19]{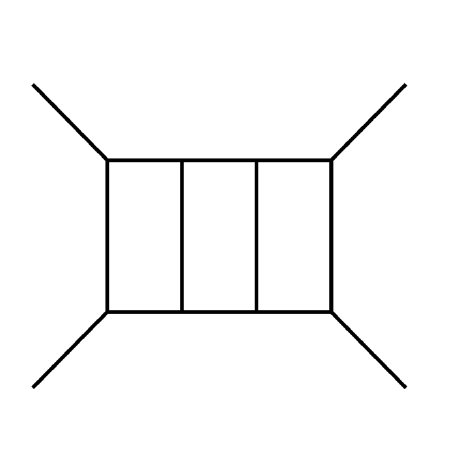}}
\subfigure[47 MIs]{\includegraphics[page=2, scale=0.19]{sections/Diagrams.pdf}}
\subfigure[53 MIs]{\includegraphics[page=3, scale=0.19]{sections/Diagrams.pdf}}
\subfigure[28 MIs]{\includegraphics[page=4, scale=0.19]{sections/Diagrams.pdf}}
\subfigure[41 MIs]{\includegraphics[page=5, scale=0.19]{sections/Diagrams.pdf}}
\subfigure[62 MIs]{\includegraphics[page=6, scale=0.19]{sections/Diagrams.pdf}}
\subfigure[87 MIs]{\includegraphics[page=7, scale=0.19]{sections/Diagrams.pdf}}
\subfigure[76 MIs]{\includegraphics[page=8, scale=0.19]{sections/Diagrams.pdf}}
\subfigure[113 MIs]{\includegraphics[page=9, scale=0.19]{sections/Diagrams.pdf}}
    \caption{The 9 independent three-loop topologies for $2\to2$ processes, and the number of master integrals (MIs), as seen in \cite{Henn:2020lye}.
    }
    \label{fig:topologies}
\end{figure}

Symmetry relations, obtained using \textsc{LiteRed}~\cite{Lee:2012cn}, were exploited to further reduce the set of independent integrals. A total of 54 independent topologies were identified out of an initial set of 158, all of which can be generated from the nine representative topologies shown in Fig.~\ref{fig:topologies} by permutations of the external momenta. The six distinct kinematic configurations arising from these permutations are reported in the first column of Table~\ref{tab:Kinematic_Regions}.

The amplitude is then decomposed into a canonical, 
$\epsilon$-factorised basis~\cite{Henn:2013pwa} consisting of 533 master integrals \cite{Henn:2020lye}, together with their five additional permutations.

\subsection{Master integrals and analytic continuation}

\subsubsection{Master integrals}
The needed master integrals for three-loop four-point massless scattering amplitudes were previously calculated in the unphysical Riemann sheet in~\cite{Henn:2020lye} in terms of generalised polylogarithms (GPLs)~\cite{Remiddi:1999ew,Goncharov:1998kja} up to transcendental weight eight, $w=8$. 
For the present calculation, we require
the evaluation of  Feynman integrals on the physical Riemann sheet. This is achieved by taking the complex conjugates of the boundary constants provided in Ref.~\cite{Henn:2020lye}.
This feature was already noted in Ref.~\cite{Caola:2020dfu}, and is further supported by numerical cross-checks we performed using \texttt{AMFlow}~\cite{Liu:2017jxz,Liu:2022chg}.

The finite remainders of the considered processes in QED involve GPLs up to $w=6$.
GPLs at weight $w=n$ are defined recursively as iterated integrals,
\begin{equation}
    G(a_{1},\dots,a_{n};x) = 
    \int_{0}^{x}\frac{dy}{y-a_{1}}G(a_{2},\dots,a_{n};y)\, ,
\end{equation}
which, for this computation, are functions of the dimensionless variable $x = -t/s$ and of the indices 
$a_i \in \left\{0,1\right\}$ that correspond to the logarithmic letters
$\{s,t,u\}$, once the dependence on $s$ has been factored out~\eqref{eq:int_I}.

\subsubsection{Analytic Continuation 
\label{app:Analytic Continuation}}

\begin{table}[t]
\begin{tabular}{cc}
\toprule
{Momenta configuration\phantom{ }} & 
{Kinematic region}     \\ 
\midrule
$(p_1,p_2,p_3,p_4)$                                  & \multirow{ 2}{*}{{$s>0 \quad t,u<0$}} \\ 
 $(p_1,p_2,p_4,p_3)$                                  & 
 \\ 
\midrule
 $(p_1,p_3,p_4,p_2)$                                  & \multirow{ 2}{*}{{$u>0 \quad s,t<0$}} \\ 
 $(p_1,p_3,p_4,p_2)$                                  &                       \\ \midrule
 $(p_1,p_4,p_2,p_3)$                                  & \multirow{ 2}{*}{{$t>0 \quad s,u<0$}} \\ 
 $(p_1,p_4,p_3,p_2)$                                  &                       \\ \bottomrule
\end{tabular}
\caption{The kinematic regions required for the corresponding momenta configurations.}
\label{tab:Kinematic_Regions}
\end{table}

The analytic expressions for the master integrals in the momenta configuration $(p_1,p_2,p_3,p_4)$, referred to as configuration 1,
are constructed by applying complex conjugation to the boundary constants given in Ref.~\cite{Henn:2020lye}.
Deriving analytic expressions for the remaining five configurations is non-trivial, as it requires analytic continuation across different kinematic regions.

The canonical system of differential equations is first taken from configuration 1 and then mapped to the remaining momenta configurations through the appropriate crossing of kinematics, yielding the corresponding differential equations.

Boundary conditions are required in each kinematic region associated with the remaining configurations. These are obtained by evaluating the master integrals in configuration 1 using \textsc{GiNaC}~\cite{Bauer:2000cp} through {\sc PolyLogTools}~\cite{Duhr:2019tlz}. The resulting boundary values are then analytically continued to the relevant kinematic regions using 
\textsc{DiffExp}~\cite{Hidding:2020ytt}. This procedure provides consistent boundary conditions for all momenta configurations.

Finally, the differential equations are solved and the integration constants are numerically matched to multiple zeta values using the PSLQ algorithm~\cite{PSLQ}, resulting in fully analytic expressions for the master integrals in all momenta configurations.

\subsection{Ultraviolet renormalisation and Infrared subtraction}

After IBP decomposition, the bare form factors are expressed in terms of a subset of 533 master integrals $M_i$, whose families are shown in Fig.~\ref{fig:topologies}, for each of the configurations of the external momenta listed in Tab.~\ref{tab:Kinematic_Regions}.
Once all Feynman integrals in the different momenta configuration are integrated wrt the same variable $x$, the coefficients of GPLs and of the associated transcendental constants are reconstructed using analytic reconstruction from numerical evaluations over finite fields~\cite{vonManteuffel:2014ixa,Peraro:2016wsq} within the \textsc{FiniteFlow} framework~\cite{Peraro:2019svx}.

The Laurent series of the expansion of the form factors around $\varepsilon = 0$  
contain poles in $\varepsilon$ that start at the sixth order, 
originating from both ultraviolet (UV) and infrared (IR) divergences.

The UV singularities are removed through renormalisation, performed in the $\overline{\text{MS}}$ scheme. 
Accordingly, the bare coupling constant $\alpha_{B}$ is replaced with its renormalised counterpart $\alpha$ using the relation:
\begin{equation}
\begin{aligned}
& \alpha_{B}\mu_{0}^{2\varepsilon} e^{-\varepsilon\gamma_{E}}
 = \\
& \qquad \alpha \mu^{2\varepsilon}
\bigg[
1 - \frac{\beta_{0}}{\varepsilon}\!\left(\frac{\alpha}{2\pi}\right)
+ \left(\frac{\beta_{0}^{2}}{\varepsilon^{2}}
       - \frac{\beta_{1}}{2\varepsilon}\right)
  \!\left(\frac{\alpha}{2\pi}\right)^{2}
\\
& \qquad
+ \left(\frac{\beta_{0}^{3}}{\varepsilon^{2}}
       - \frac{7\beta_{0}\beta_{1}}{6\varepsilon^{2}}
       + \frac{\beta_{2}}{3\varepsilon}\right)
  \!\left(\frac{\alpha}{2\pi}\right)^{3}
+ \mathcal{O}(\alpha^{4})
\bigg] ,
\end{aligned}
\end{equation}
where $\mu$ is the renormalisation scale and $\beta_i$ are the coefficients of the QED $\beta$-function~\cite{Herzog:2017ohr,Baikov:2016tgj}. 
These coefficients are obtained by abelianising the corresponding QCD expressions ($T_f=1, C_F=1, C_A=0$) and are presented explicitly in this work, with full expressions given in App.~\ref{app:IR}.

To renormalise the three-loop form factors, and consequently the  amplitude, it is necessary to combine the bare contributions from all perturbative orders, ranging from tree level up to three loops. At the amplitude level, this procedure amounts to
\begin{equation}
    \begin{aligned}
 \mathcal{A}^{(3)} = & \mathcal{A}_B^{(3)}-\frac{3\beta_0}{\varepsilon}\mathcal{A}_B^{(2)}+\left(\frac{3\beta_0^2 -\varepsilon\beta_1}{\varepsilon^2}\right)\mathcal{A}_B^{(1)}\\
 & 
 -\left(\frac{6\beta_0^3-7\varepsilon\beta_1\beta_0+2\varepsilon^2\beta_2}{6\varepsilon^3}\right)\mathcal{A}_B^{(0)}\,.
    \end{aligned}
\end{equation}
The form factors and the amplitudes for tree-level $({\cal A}_B^{(0)})$, one-loop $({\cal A}_B^{(1)})$ up to $\varepsilon^4$ and two-loop $({\cal A}_B^{(2)})$ up to $\varepsilon^2$ have been computed in \cite{Dave:2024ewl}.

After UV renormalisation, the remaining IR singularities, appearing as residual poles in $\varepsilon$, are removed by an IR subtraction procedure. The finite remainder ${\cal R}^{(3)}$ of the three-loop amplitude ${\cal A}^{(3)}$ then follows from the IR subtraction formula:
\begin{equation}
    \mathcal{R}^{(3)} = \mathcal{A}^{(3)}-\mathcal{I}^{(1)}\mathcal{A}^{(2)}-\mathcal{I}^{(2)}\mathcal{A}^{(1)}-\mathcal{I}^{(3)}\mathcal{A}^{(0)}\,,
    \label{eq:finite_remainder}
\end{equation}
which combines the renormalised amplitudes up to three-loop order with the IR subtraction operators  $\mathcal{I}^{(n)}$ for $n=1,2,3$.
The renormalised tree-, one-, and two-loop amplitudes are taken from Ref.~\cite{Dave:2024ewl}. The IR subtraction operators $\mathcal{I}^{(n)}$ are obtained by abelianising the corresponding operators originally derived within the soft-collinear effective theory (SCET) framework for the subtraction of infrared singularities in QCD~\cite{Becher:2009qa}.
Explicit expressions for $\mathcal{I}^{(n)}$ are given in App.~\ref{app:IR}.

The expressions for the finite remainder ${\cal R}^{(3)}$ of the three-loop corrections to four-fermion scattering amplitudes in QED 
contain GPLs in the variable $x$
up to transcendental weight six, and 
constitute the main novel result of this work.

\section{Results}

\begin{table*}[t]
    \begin{tabular}{llcc}
    \toprule
    Process  & ${\cal A}$ & ${\cal H}$ & ${\Phi}$\\
    \midrule 
    \multirow{ 2}{*}{$0 \to e^+e^- \mu^+ \mu^-\qquad$}
     & 
     $\mathcal{A}{(1_{e^+}^-,2_{e^-}^+,3_{\mu^+}^+,4_{\mu^-}^-)}\quad$
     & $  
            -2 \mathcal{F}_{\mu\mu,1} 
            +\left(1-x\right)\mathcal{F}_{\mu\mu,2}
            $ & $
            \quad \langle 1 4 \rangle [2 3] \quad
            $  \\
    & 
            $\mathcal{A}{(1_{e^+}^-,2_{e^-}^+,3_{\mu^+}^-,4_{\mu^-}^+)}$
            & $             
            -2 \mathcal{F}_{\mu\mu,1} -x\mathcal{F}_{\mu\mu,2}  $
            & $
            \quad \langle 1 3 \rangle [2 4] \quad
            $ \\ 
    \midrule
    \multirow{ 2}{*}{
    $0 \to e^+ \mu^- \mu^+ e^- \qquad$
    }  & 
            $\mathcal{A}{(1_{e^+}^-,2_{\mu^-}^-,3_{\mu^+}^+,4_{e^-}^+)}$
            & $
            2 \mathcal{F}_{e\mu,1} 
            -\left(1-x\right)\mathcal{F}_{e\mu,2}  
            $ 
            & $
            \quad \langle 1 2 \rangle [3 4] \quad
            $ \\ 
                & 
            $\mathcal{A}{(1_{e^+}^-,2_{\mu^-}^+,3_{\mu^+}^-,4_{e^-}^+)}$
            & $
            2 \mathcal{F}_{e\mu,1} 
            -\mathcal{F}_{e\mu,2}
            $ & $
            \quad \langle 1 3 \rangle [2 4] \quad
            $ \\ 
    \midrule
    \multirow{2}{*}{$0 \to e^+ e^- e^+ e^-$}  & 
            $\mathcal{A}{(1_{e^+}^-,2_{e^-}^+,3_{e^+}^-,4_{e^-}^-)}$
            &$ 
            -2 \mathcal{F}_{\mu\mu,1} 
            +\left(1-x\right)\mathcal{F}_{\mu\mu,2}
            $ 
            & $ \langle 1 4 \rangle [2 3] 
            $  \\ 
            & 
            $\mathcal{A}{(1_{e^+}^-,2_{e^-}^+,3_{e^+}^-,4_{e^-}^+)}$
            & $ 
            -2 \mathcal{F}_{\mu\mu,1} 
            -x\mathcal{F}_{\mu\mu,2}
            -2\mathcal{F}_{e\mu,1}+\mathcal{F}_{e\mu,2}
            $
            & $ \langle 1 3 \rangle [2 4] 
            $ \\ 
            \hline
    \end{tabular}
    \caption{Helicity amplitudes structure:
        the helicity amplitude ${\cal A}$,
        the tensor structure ${\cal H}$, and 
        the spinor prefactor ${\Phi}$ are shown, process by process 
        (for $s =1$, $x=-t$ and $1-x = -u$).
    }\label{tab:helicityamplitudes}
\end{table*}

We hereby present the result of the three-loop helicity amplitudes for the four-lepton processes:
\begin{subequations}
\label{eq:qed_processes}
\begin{align}
\label{eemumu}
e^{+}e^{-}&\rightarrow\mu^{+}\mu^{-}\,,
\\
\label{eMueMu}
e^{+}\mu^{-}&\rightarrow \mu^{+}e^{-}\,,
\\
\label{Bhahba}
e^{+}e^{-}&\rightarrow e^{+} e^{-}\,.
\end{align}
\end{subequations}

\subsection{
$e^{+}e^{-}\rightarrow\mu^{+}\mu^{-}$ and 
$e^{+}\mu^{-}\rightarrow \mu^{+}e^{-}$
}

In line with the tensor decomposition~\eqref{eq:eemmumu_tensor_decomposition},
the amplitudes of the first two processes, generically denoted by
$a\,b\to c\,d$, are decomposed as~\cite{Chen:2019wyb,Peraro:2019cjj,Peraro:2020sfm},
\begin{align}
    {\cal A} = 
      {\cal F}_{1, cd} \, {\cal T}_{1, cd} 
    + {\cal F}_{2, cd} \, {\cal T}_{2, cd} \ ,
    \label{eq:abcd_amplitudes}
\end{align}
in terms of
two tensors ${\cal T}_{1, cd}$
and ${\cal T}_{2, cd}$, 
and their corresponding form factors
${\cal F}_{1, cd}$
and ${\cal F}_{2, cd}$,
labelled by the final state particles $c$ and $d$. 
For the dimuon production process ($c=d=\mu$),
the two tensors are given in 
eq.~\eqref{eq:eemumu_Tensor_Structures}.
The tensors relevant for elastic positron-muon scattering 
($c=e, d=\mu$) follow from these expressions by crossing: 
$p_2 \leftrightarrow p_4$, corresponds to the exchange of Mandelstam invariants $s \leftrightarrow t $,
\begin{equation}
    \begin{aligned}
 & \mathcal{T}_{1,e\mu}=\overline{u}(p_{4})\gamma_{\sigma}v(p_{1})\times\overline{u}(p_{2})\gamma^{\sigma}v(p_{3})\,, \\
 & \mathcal{T}_{2,e\mu}=\overline{u}(p_{4})\fsl{p_{3}}v(p_{1})\times\overline{u}(p_{2})\fsl{p_{1}}v(p_{3})\,.
\end{aligned}
\end{equation}

The tensors 
$\mathcal{T}_{\mu\mu,i}$ and 
$\mathcal{T}_{e\mu,i}$  with $i=1,2$ depend implicitly on the helicities of the external lepton spinors. As a result, for fixed helicity assignments the corresponding helicity amplitudes in eq.~\eqref{eq:abcd_amplitudes}
admit a factorised form,
\begin{align}
&\mathcal{A}{(
1_a^{\lambda_1},
2_b^{\lambda_2},
3_c^{\lambda_3},
4_d^{\lambda_4}
)}
= \mathcal{H} \, \Phi \,.
\label{eq:helamp_factorization}
\end{align}
where, ${\cal H}$ is a linear combination of the form factors, while $\Phi$ depends solely on spinor products~\cite{Dixon:1996wi}.
For brevity, the helicity dependence of ${\cal H}$ and $\Phi$ is left implicit.

Within the massless-lepton approximation, only two independent non-vanishing helicity configurations contribute for each scattering process; the remaining ones follow from parity and/or charge conjugation.
The independent helicity configurations and the corresponding expressions for ${\cal H}$ and $\Phi$ are listed in the first two rows of Tab.~\ref{tab:helicityamplitudes} for dimuon production in electron–positron annihilation and for elastic positron–muon scattering, respectively.

\subsection{$e^{+}e^{-}\rightarrow e^{+}e^{-}$}
The amplitudes of the Bhabha scattering
can be obtained by an appropriate combination of the amplitudes of the two processes discussed above. 
In fact, the $s$- and $t$-channel diagrams of $e^{+}e^{-}\rightarrow e^{+}e^{-}$ coincide with those of $e^+e^- \to \mu^+\mu^-$ and 
$e^+ \mu^-\to \mu^+ e^-$, respectively
\begin{align}
   {\cal A}_{e^+ e^- \to e^+ e^-} 
  & = 
   {\cal A}_{e^+ e^- \to \mu^+ \mu^-}
   - {\cal A}_{e^+ \mu^- \to \mu^+ e^-} \ , 
   \label{eq:bhabha_vs_eemumu_emuemu}
\end{align}
upon identifying the muon with the electron as equivalent massless particles.
The relative minus sign arises from Fermi–Dirac statistics and reflects the antisymmetry of the amplitude under the exchange of identical fermions, in accordance with the Pauli principle.

Therefore, the helicity amplitudes for Bhabha scattering also admit the factorised representation of eq.~\eqref{eq:helamp_factorization} in terms of ${\cal H}$ and ${\Phi}$.
Because of the relation in eq.~\eqref{eq:bhabha_vs_eemumu_emuemu}, the quantity ${\cal H}$ is given by a linear combination of the four form factors associated with the two previously considered processes, 
${\cal F}_{i,\mu\mu}$ and ${\cal F}_{i,\mu e}$ 
with $i=1,2$.

The helicity amplitudes for the two independent non-vanishing helicity configurations of Bhabha scattering, together with the corresponding expressions for ${\cal H}$ and ${\Phi}$, are reported in the third row of Tab.~\ref{tab:helicityamplitudes}.

{\it Format}. 
Given their size, the three-loop amplitude results are provided in electronic form and are available at Ref.~\cite{dave_2026_17855694}. Specifically, this repository contains the UV-renormalised and IR-finite expressions for the form factors of the processes in eqs.~\eqref{eemumu} and \eqref{eMueMu}, as well as the UV-renormalised and IR-finite helicity amplitudes for all processes listed in eq.~\eqref{eq:qed_processes}. 
All results are expressed in terms of GPLs up to transcendental weight six with argument~$x$.

\subsection{Checks}

{\it Finiteness.} 
The finiteness of the combination appearing on the right-hand side of eq.~\eqref{eq:finite_remainder} in the limit $\varepsilon \to 0$ provides a stringent validation of the entire calculation, as well as of the non-trivial methods and software developed to carry it out.

{\it QED vs QCD.}
Additional checks are performed by isolating the Abelian contributions of QCD diagrams within the known three-loop QCD corrections to four-quark scattering amplitudes~\cite{Caola:2021rqz}, 
which directly map onto their QED counterparts.

For each process, the three-loop QED amplitudes
$\mathcal{A}^{(3)}$  can be decomposed in terms of gauge invariant coefficients proportional to the number $N_f$ of closed fermion loops, 
\begin{align}
    \mathcal{A}^{(3)}= 
    \sum_{i=0}^3 
    \mathcal{A}_{[i]}^{(3)} \, N_f^i \ .
\end{align} 

Therefore,  by casting the corresponding three-loop QCD amplitudes 
$\mathcal{M}^{(3)}$ \cite{Caola:2021rqz} 
in terms of a double sum, with coefficients proportional to $N_f$ and 
to the number of colours $N_c$,
\begin{align}
    \mathcal{M}^{(3)}=
    \sum_{i= 0}^{3} 
    \sum_{j=-4}^{2} 
    \mathcal{M}_{[i,j]}^{(3)} \, N_f^i \, N_c^j
    \,,
\end{align} 
we verified that 
$
{\cal M}_{[i,j]}^{(3)} = \frac{(-1)^i}{2} \, 
{\cal A}_{[i]}^{(3)} 
$,
for $i=0,2,3$ and $j=-4+i$,
since $\mathcal{M}_{[i,j]}^{(3)}$, in these cases, 
receives contributions exclusively from combinations of Abelian diagrams (with the quark colour factors set to unity), each entering with the same numerical coefficient.
In contrast, due to the colour algebra, we verified that 
$\mathcal{M}_{[1,j]}^{(3)}$ receives contributions from two distinct sets of Abelian diagrams that appear in the QED counterpart $\mathcal{A}_{[1]}^{(3)}$.
Specifically, 
$\mathcal{M}_{[1,j]}^{(3)} = 
-\frac{1}{2}\mathcal{A}_{[1,1]}^{(3)} 
-\frac{3}{2}\mathcal{A}_{[1,2]}^{(3)}
\, ,$
where $\mathcal{A}_{[1,k]}^{(3)}$ for $k=1,2$ 
are two components of $\mathcal{A}_{[1]}^{(3)}$,
such that $
\mathcal{A}_{[1]}^{(3)}
=
\mathcal{A}_{[1,1]}^{(3)}
+
\mathcal{A}_{[1,2]}^{(3)}
$.
\section{Conclusion}

In this work, we presented the complete analytic three-loop corrections to four-fermion scattering amplitudes in massless QED. We focused on a process involving distinct fermion flavours, $e^+e^- \to \mu^+\mu^-$, and exploited crossing symmetries to obtain 
the amplitudes for $e \mu$ scattering and for 
 Bhabha scattering,  
$e^+e^- \to e^+e^-$.

The three-loop scattering amplitudes were decomposed into four-dimensional Lorentz-invariant tensor structures, and the associated Feynman integrals were regulated using dimensional regularisation. To overcome the complexity of the three-loop calculation, we employed an in-house graph mapping algorithm in combination with an integration-by-parts reduction framework that makes use of spanning cuts, syzygy-based identities, and an optimised seeding strategy, leading to a substantial reduction in the size of the intermediate systems and in the overall computational cost.

The required master integrals for all kinematic configurations were obtained from the known expressions of a single reference configuration via exchange of Mandelstam variables within the canonical differential equations alongside high-precision numerical fitting of boundary conditions, after analytic continuation, to determine the boundary constants.
The final independent helicity amplitudes are given in terms of generalised polylogarithms up to transcendental weight six of a single dimensionless variable.

These results represent an important milestone in the computation of QED scattering processes at three-loop order in the high-energy regime, where fermion masses can be consistently neglected.

~

\section*{Acknowledgements}

We thank Manoj Mandal for collaboration at early stages.
We thank Mao Zeng for granting access to his private Mathematica–Singular interface and to his implementation of the algorithm for identifying non-vanishing sectors.
We thank Andreas Von Manteuffel for a careful reading of and comments on the manuscript.
T.D. and W.J.T. thank J\'er\'emy Paltrinieri, Pau Petit Ros\`as and Mattia Pozzoli  for collaboration on closely related projects, and Giulio Gambuti, Lorenzo Tancredi and Yannick Ulrich for useful discussions.
We acknowledge the stimulating interactions within the theory initiative of the 
{\sc MUonE} CERN experiment, triggering the motivation for this project.
G.C., P.M., J.R., S.S.
gratefully acknowledge the stimulating scientific environments of “Domoschool 2025” and “Loop-The-Loop 2025”, which provided valuable venues for discussions and development of some of the novel algorithms  used in this work. 
G.C., P.M., J.R., S.S. acknowledge the partial  support of the 
{\it Amplitudes} INFN research initiative.
The work of T.D. and W.J.T. is supported by the Leverhulme Trust, LIP-2021-014.

\appendix
\begin{widetext}
\section{IR subtraction operators and $\beta$ functions}
\label{app:IR}

The $\beta$-functions that are required for both UV renormalisation and IR subtraction are defined as,
\begin{align}
\frac{d\alpha_s}{d\log(\mu)}&=\beta(\alpha_s)-2\varepsilon\alpha_s\,, && \beta(\alpha_s) = -2\alpha_s \sum_{n=0}^{\infty} \beta_n \left(\frac{\alpha_s}{2\pi}\right)^{n+1}\,,
\end{align}
whose coefficients $\beta_i$, up to the second order in $\alpha_s$, read as,
\begin{equation}
    \beta_{0} = -\frac{2N_{f}}{3}, \quad \beta_{1} = -N_{f}, \quad \beta_{2} = \frac{N_{f}}{4}+\frac{11N_{f}^{2}}{18}\,.
\end{equation}
The IR subtraction operators $\mathcal{I}^{(n)}$,  for the $n$-th order corrections, 
are defined as 
\begin{equation}
    \begin{aligned}
    \mathcal{I}^{(1)} &= \mathcal{Z}^{(1)}\,,\\
    \mathcal{I}^{(2)} &= \mathcal{Z}^{(2)}-(\mathcal{Z}^{(1)})^{2}\,,\\
    \mathcal{I}^{(3)} &= \mathcal{Z}^{(3)}-2\mathcal{Z}^{(2)}\mathcal{Z}^{(1)}-(\mathcal{Z}^{(1)})^{3}\,,
    \end{aligned}
\end{equation}
in terms of the coefficients $\mathcal{Z}^{(n)}$   of the series expansion of 
the multiplicative colour-space operator $\mathcal{Z}$~\cite{Ahrens:2012qz,Becher:2009cu,Almelid:2015jia}, 
\begin{equation}
    \mathcal{Z}(\varepsilon,\{p\},\mu) = \mathbb{P}\exp\left[\int_{\mu}^{\infty}\frac{d\mu'}{\mu'}\Gamma(\{p\},\mu')\right]\,,
\end{equation}
with $\Gamma$ being the anomalous dimension matrix.
Explicitly, they read as,
\begin{equation}
    \begin{aligned}
        &\mathcal{Z}^{(1)}= \frac{{\Gamma_0^{\prime}}^2}{4\varepsilon^2}+\frac{\Gamma_0}{2\varepsilon}\,, &&\\
&\mathcal{Z}^{(2)}=  \frac{{\Gamma_0^{\prime}}^2}{32\varepsilon^4}+\frac{\Gamma_0}{8\varepsilon^3}\left(\Gamma_0-\frac{3}{2}\beta_0 \right)+\frac{1}{4\varepsilon^2}\left(-\beta_0\Gamma_0+\frac{{\Gamma_0}^2}{2}+\frac{\Gamma_1^{\prime}}{4} \right)+\frac{\Gamma_1}{4\varepsilon}\,,&&\\
&\mathcal{Z}^{(3)}=  \frac{{\Gamma_0^{\prime}}^3}{384\varepsilon^6}+\frac{{\Gamma_0^{\prime}}^2}{64\varepsilon^5}\left(\Gamma_0-3\beta_0 \right)+\frac{\Gamma_0^{\prime}}{32\varepsilon^4}\left(\Gamma_0-\frac{4}{3}\beta_0 \right)\left(\Gamma_0-\frac{11}{3}\beta_0 \right)+\frac{\Gamma_0^{\prime}\Gamma_1^{\prime}}{64\varepsilon^4}&&\\
& \hspace{1cm}+\frac{\Gamma_0}{48\varepsilon^3}\left(\Gamma_0-2\beta_0 \right)\left(\Gamma_0-4\beta_0 \right)+\frac{\Gamma_0^{\prime}}{16\varepsilon^3}\left(\Gamma_1-\frac{16}{9}\beta_1 \right)+\frac{\Gamma_1^{\prime}}{32\varepsilon^3}\left(\Gamma_0-\frac{20}{9}\beta_0 \right)&&\\
& \hspace{1cm}+\frac{\Gamma_0\Gamma_1}{8\varepsilon^2}-\frac{\beta_0\Gamma_1+\beta_1\Gamma_0}{6\varepsilon^2}+\frac{\Gamma_2^{\prime}}{36\varepsilon^2}+\frac{\Gamma_2}{6\varepsilon} \ .&&
    \end{aligned}
\end{equation}
The elements of the anomalous dimension matrix $\Gamma$ are defined as,
\begin{equation}
\Gamma_n = -2\gamma_n^{\text{cusp}}\left(L_{34}+L_{23}-L_{24}\right)+4\gamma_n^q\,,
\end{equation}
and
\begin{equation}
\Gamma_n^{'} = \frac{\partial\Gamma_l}{\partial \log(\mu)}= -4\gamma_n^{\text{cusp}}\,,
\end{equation}
where 
\begin{equation}
L_{ij}= \log\left(\frac{\mu^2}{-s_{ij}-i0}\right)\,.
\end{equation}
Below, we list the anomalous dimension coefficients for the particles in our processes. We obtained these values considering the coefficients of 
$\Big({\alpha_{\text S}^{(n_l)}}/{2\pi}\Big)^{n+1}$ in the expansions given in \cite{Moch:2004pa,Vogt:2004mw,Moch:2005tm,Moch:2005id}, upon abelianisation, namely by setting $C_A=0$, $C_F=1$ and $T_F=1$.
For the cusp anomalous dimension, we find,
\begin{equation}
\begin{aligned}
\label{cuspAD}
 &\gamma_0^{\text{cusp}}=2\,, &&\\
&\gamma_1^{\text{cusp}}=-\frac{20}{9}N_f\,,&&\\
&\gamma_2^{\text{cusp}}=\left(-\frac{55}{6} +8\zeta_3\right)N_f -\frac{8}{27}{N_f}^2\,.&&
\end{aligned}
\end{equation}
Whereas for the quark anomalous dimension we find;
\begin{equation}
\begin{aligned}
\label{quarkAD}
 &\gamma_0^{q}=-\frac{3}{2}\,, &&\\
&\gamma_1^{q}=\left(\frac{65}{54}+\frac{\pi^2}{6}\right)N_f+\left(\frac{\pi^2}{2}-6\zeta_3-\frac{3}{8}\right)\,,&&\\
&\gamma_2^{q}=\left(\frac{29}{16}+\frac{3\pi^2}{8}+\frac{\pi^2}{5}+\frac{17\zeta_3}{2}+\frac{2\pi^2 \zeta_3}{3}+30\zeta_5\right)+\left(\frac{2953}{216}-\frac{13\pi^2}{36}-\frac{7\pi^4}{54}+\frac{64\zeta_3}{9}\right)N_f&&\\
&\hspace{1cm}+\left(\frac{2417}{1458}-\frac{5\pi^2}{27}-\frac{4\zeta_3}{27}\right){N_f}^2\,.&&
\end{aligned}
\end{equation}

\end{widetext}

\mbox{~}
\newpage
\bibliographystyle{JHEP}
\bibliography{biblio}
\end{document}